\documentclass[onecolumn]{aa}
\usepackage{amsmath}
\usepackage{amssymb}
\usepackage{upgreek}
\usepackage[varg]{txfonts}
\usepackage{epsfig,graphicx}
\usepackage{xcolor}
\usepackage[colorlinks=true, linkcolor=blue, citecolor=blue, urlcolor=blue]{hyperref}
\usepackage{natbib}
\bibpunct{(}{)}{;}{a}{}{,} 
\newcommand{\Rb}{R_{\text{b}}}
\newcommand{\Reff}{R_{\text{e}}}
\newcommand{\Ib}{I_{\text{b}}}
\newcommand{\txd}{{\text{d}}}

\begin{document}

\title{The Nuker model for galactic nuclei}

\author{Maarten Baes\inst{\ref{UGent}}}

\institute{
Sterrenkundig Observatorium, Universiteit Gent, Krijgslaan 281 S9, 9000 Gent, Belgium
\label{UGent}
}

\abstract{The Nuker profile, characterised by an inner and outer power-law profile smoothly merged around a break radius, is a very popular model to describe the surface brightness profile of galactic nuclei. A disadvantage of this model for dynamical studies is that the spatial density distribution that corresponds to this surface brightness profile cannot be written in terms of elementary or regular special functions. We derive a compact and elegant analytical expression for the density of the Nuker model, based the Mellin integral transform method. We use this expression to discuss the general behaviour and asymptotic expansion of the density. We also discuss the special subclass of Nuker models with an infinitely sharp break, and demonstrate that these models are always characterised by non-monotonous and hence unphysical density profile. We extend our study to the dynamical structure of spherical isotropic galactic nuclei with a Nuker surface brightness profile. Based on this analysis, we extend and refine the classification of spherical isotropic galactic nuclei introduced by Tremaine et al.\ (1994). We demonstrate that both the inner density slope and the sharpness of the break between the inner and outer profiles critically determine the consistency and stability of the Nuker models. 
}


\maketitle

\section{Introduction}

The first high-resolution observations of galactic nuclei with the Hubble Space Telescope showed that their surface brightness profiles are generally characterised by power-law behaviour at small radii \citep{1991ApJ...369L..41L, 1992AJ....103..703L, 1993AJ....106.1371C, 1994AJ....108.1598F}. \citet{1995AJ....110.2622L} proposed a relatively simple model for the surface brightness profile of galactic nuclei,
\begin{equation}
I(R)
=
2^{\frac{\beta-\gamma}{\alpha}}\,\Ib \left(\frac{R}{\Rb}\right)^{-\gamma}
\left[ 1+\left(\frac{R}{\Rb}\right)^\alpha \right]^{\frac{\gamma-\beta}{\alpha}}.
\label{Nuker}
\end{equation}
In this formula, $R$ is the circular radius on the plane of the sky, $\Rb$ is the break radius that indicates the transition between the inner and the outer profile, $\Ib=I(\Rb)$ is the surface brightness at the break radius, $\beta$ and $\gamma$ correspond to the negative logarithmic surface brightness slopes at large and small radii respectively, and $\alpha$ is a parameter that sets the width of the transition between the inner and outer profiles. This model has become generally known in the extragalactic community as the Nuker model.

The availability of a simple model that accurately describes the surface brightness profile of galactic nuclei is important for many reasons. Numerous studies have used this model to parameterise the surface brightness profiles of the central regions of galaxies \citep[e.g.,][]{1996AJ....111.1889B, 2000ApJS..128...85Q, 2001AJ....121.2431R, 2003AJ....125..478L, 2005A&A...439..487D, 2005AJ....129.2138L, 2007ApJ...664..226L, 2019A&A...622A..78D}. Based on this model, a bimodal distribution in the central structure of luminous elliptical galaxies has been discovered, with most galaxies either power-law systems with steep cusps, or core systems with shallow cusps interior to a steeper envelope brightness distribution \citep{1995AJ....110.2622L, 2007ApJ...664..226L, 1997AJ....114.1771F}. 

One disadvantage of the Nuker model, particularly for theoretical studies of the structure and dynamics of galactic nuclei, is that the spatial density distribution that corresponds to the surface brightness profile (\ref{Nuker}) is not easy to obtain, even in the simplifying assumption of spherical symmetry. Indeed, inserting the surface brightness profile into the standard de-projection formula yields an integral that cannot be evaluated in terms in elementary functions. This is unfortunate, as the spatial density is probably the most fundamental property for dynamical studies \citep{1986PhR...133..217D}. Most popular families of dynamical models have a simple analytical expression for the spatial mass density as a starting point \citep[e.g.,][]{1993MNRAS.265..250D, 1994AJ....107..634T, 1996MNRAS.278..488Z, 2005MNRAS.360..492E}.

One way to deal with this problem is to use a model that approximates the Nuker model, but in which the density has a simple analytical expression. In particular, \citet{1996MNRAS.278..488Z} explored a family of models in which the spatial density profile rather than the surface brightness profile is characterised by a double power-law model, very similar to Eq.~(\ref{Nuker}). This general family of models, also called the generalised NFW models, is sufficiently simple that the potential and many other interesting dynamical properties can be expressed in terms of elementary or standard special functions. In a follow-up paper, \citet{1997MNRAS.287..525Z} provided a recipe to approach the Nuker model with one of the generalised NFW models, by matching the surface brightness profiles in a least squares sense. 

\citet{2005MNRAS.356.1403R} took this approach one step further, and used a linear combination of simple models to approach the Nuker model. More specifically, they used a quadratic programming technique \citep{1989ApJ...343..113D} to minimise the difference of the Nuker surface brightness profile and the one corresponding to a linear combination of generalised NFW models. They showed that only a few components are necessary to obtain a good fit. Still, this approach is not guaranteed to have the correct behaviour at all radii, in particular around the transition region between the inner and outer profile.

In this paper, we use a different and direct approach to study the structure and dynamics of the spherical systems characterised by the Nuker surface brightness profile. We make use of an advanced analytical technique based on the Mellin transform method \citep{marichev1983handbook, fikioris2007mellin}. This approach has the advantage that it yields exact analytical expressions for the spatial density and related properties, which allows for a detailed study of the properties and asymptotic behaviour of the model. A similar approach has been adopted to explore the structure and dynamical properties of spherical models with a S\'ersic surface brightness profile \citep{2011A&A...525A.136B, 2011A&A...534A..69B, 2019A&A...626A.110B, 2019A&A...630A.113B}.

This paper is organised as follows. In Sec.\ {\ref{Density.sec}} we derive an analytical expression for the spatial density profile of the Nuker model using the Mellin integral transform method, we discuss the general behaviour and asymptotic expansion, and we look in detail at the special subclass of models with an infinitely sharp break. In Sec.\ {\ref{Dynamics.sec}} we discuss some important dynamical properties of the Nuker model, including the potential, velocity dispersion and phase-space distribution function. In Sec.\ {\ref{Sersic.sec}} we discuss a special subset of Nuker models, namely the family of S\'ersic models. In Sec.\ {\ref{Discussion.sec}} we discuss and summarise our results, where we mainly use the family of Nuker models to refine the classification of galactic nuclei proposed by \citet{1994AJ....107..634T}.


\section{Spatial density of the Nuker model}
\label{Density.sec}

\subsection{Setting the scene}

The Nuker model is characterised by the surface brightness profile (\ref{Nuker}). As the the break radius and the surface brightness at that radius are just scaling parameters, we can consider this family of models a three-parameter family, characterised by the triplet $(\alpha,\beta,\gamma)$. In fact, in the remainder of this paper, in order to simplify the expressions, we use normalised units, with $G = M = L = \Rb = 1$, where $G$ is the gravitational constant, $M$ the total mass, and $L$ the total luminosity. The total luminosity of the Nuker model is found by integrating the surface brightness profile over the plane of the sky,
\begin{equation}
L = 2\pi\int_0^\infty I(R)\,R\,{\text{d}}R.
\end{equation}
If $\alpha>0$, $\beta>2$ and $\gamma<2$, we find a finite luminosity
\begin{equation}
L
=
\frac{2^{\frac{\beta-\gamma}{\alpha}+1}\,\pi}{\alpha}\,
\frac{\Gamma\left(\frac{\beta-2}{\alpha}\right) \Gamma\left(\frac{2-\gamma}{\alpha} \right)}
{\Gamma\left(\frac{\beta-\gamma}{\alpha}\right)}\,
\Ib\,\Rb^2.
\label{L}
\end{equation}
Using this expression and the convention of normalised units, we can rewrite the surface brightness profile of the Nuker model as
\begin{equation}
I(R) = 
\frac{\alpha}{2\pi}\,
\frac{\Gamma\left(\frac{\beta-\gamma}{\alpha}\right)}{\Gamma\left(\frac{\beta-2}{\alpha}\right) 
\Gamma\left(\frac{2-\gamma}{\alpha} \right)}\,
R^{-\gamma}
\left( 1+R^\alpha \right)^{\frac{\gamma-\beta}{\alpha}}.
\label{sb}
\end{equation}
We limit the parameters to the ranges indicated above, with the exception that we only consider positive values of $\gamma$ to avoid unrealistically increasing surface density profiles at small radii. Hence, in the remainder of this paper, we assume $\alpha>0$, $\beta>2$ and $0\leqslant\gamma<2$.


\subsection{Analytical expression for the density}

Assuming a constant mass-to-light-ratio, the density profile $\rho(r)$ can be found from the surface brightness profile $I(R)$ through the standard deprojection formula \citep[e.g.,][]{2008gady.book.....B},
\begin{equation}
\rho(r)
=
-\frac{1}{\pi}\int_r^\infty \frac{{\text{d}}I(R)}{\txd R}\,\frac{\txd R}{\sqrt{R^2-r^2}}.
\label{deprojection}
\end{equation}
This gives
\begin{equation}
\rho(r)
=
\frac{\alpha}{2\pi^2}\,
\frac{\Gamma\left(\frac{\beta-\gamma}{\alpha}\right)}{\Gamma\left(\frac{\beta-2}{\alpha}\right) 
\Gamma\left(\frac{2-\gamma}{\alpha} \right)}\,
\int_r^\infty
\frac{R^{-1-\gamma}\left(1+R^\alpha\right)^{\frac{\gamma-\beta}{\alpha}-1}\,
(\beta R^\alpha+\gamma)\,\txd R}
{\sqrt{R^2-r^2}}.
\label{rho-integral}
\end{equation}
The integral in this expression can not be readily evaluated using standard methods. To find an analytical expression, we follow a method known as the Mellin transform method \citep{marichev1983handbook, fikioris2007mellin}. While the Mellin transform is a well-known integral transform used in a variety of fields as statistics \citep{Epstein1948, fox_1957}, analytic number theory \citep{COFFEY2015507}, signal and image analysis \citep{Casasent1976258, WU200375}, or stellar dynamics \citep{1986PhR...133..217D}, its application to analytically evaluate integrals is less well-known. The method is surprisingly flexible and powerful, however, and it often yields closed-form expressions that are very difficult to come up with using other methods. We start by rewriting expression (\ref{rho-integral}) in the form
\begin{equation}
\rho(r)
=
\int_0^\infty f_1(R)\,f_2\left(\frac{1}{R}\right) \frac{{\text{d}}R}{R},  
\label{Mellinconvolution}
\end{equation}
with
\begin{equation}
f_1(R)
=
\frac{\alpha}{2\pi^2}\,
\frac{\Gamma\left(\frac{\beta-\gamma}{\alpha}\right)}{\Gamma\left(\frac{\beta-2}{\alpha}\right) 
\Gamma\left(\frac{2-\gamma}{\alpha} \right)}\,
R^{-\gamma}\,(1+R^\alpha)^{\frac{\gamma-\beta}{\alpha}-1}\,
(\beta R^\alpha+\gamma),
\end{equation}
and
\begin{equation}
f_2(R)
=
\begin{cases}
\;\dfrac{R}{\sqrt{1-r^2R^2}}
&\qquad\text{if }0\leqslant R < \dfrac{1}{r}, \\
\; 0
&\qquad\text{else}.
\end{cases}
\end{equation}
Expression~(\ref{Mellinconvolution}) is a Mellin convolution of the two functions $f_1$ and $f_2$. Similarly to the well-known Fourier convolution theorem, the Mellin transform of a Mellin convolution is equal to the product of the Mellin transforms of the two original functions. Specifically, it implies that we can evaluate the expression~(\ref{Mellinconvolution}) as the inverse Mellin transform of the product of the Mellin transforms of $f_1$ and $f_2$. Explicitly, we find
\begin{equation}
\rho(r)
=
\frac{1}{2\pi i}
\int_{\mathcal{L}} {\mathfrak{M}}_{f_1}(u)\, {\mathfrak{M}}_{f_2}(u)\, \txd u,
\end{equation}
where ${\mathfrak{M}}_{f}(u)$ denotes the Mellin transform of a function $f$, and the contour integration is along a vertical path $\mathcal{L}$ in the complex plane \citep[for more details on Mellin transforms and their inverse, see][]{fikioris2007mellin}. The Mellin transforms of the functions $f_1$ and $f_2$ can be calculated exactly,
\begin{gather}
{\mathfrak{M}}_{f_1}(u)
=
\frac{1}{2\pi^2}\,
\frac{\Gamma\left(\tfrac{\beta-u}{\alpha}\right)\,\Gamma\left(\tfrac{u-\gamma}{\alpha}\right)}
{\Gamma\left(\frac{\beta-2}{\alpha}\right)\Gamma\left(\frac{2-\gamma}{\alpha} \right)}\,
u,
\\
{\mathfrak{M}}_{f_2}(u) 
= 
\frac{\sqrt{\pi}\,\Gamma\left(\frac{1+u}{2}\right)}{\Gamma\left(\frac{u}{2}\right)}\,\frac{1}{u\,r^{1+u}}.
\end{gather}
This yields the following expression,
\begin{equation}
\rho(r)
=
\frac{1}{\pi^{3/2}}\,
\frac{1}{\Gamma\left(\frac{\beta-2}{\alpha}\right) 
\Gamma\left(\frac{2-\gamma}{\alpha} \right)}\,r^{-1}
\frac{1}{2\pi i}
\int_{\mathcal{L}} 
\frac{  
    \Gamma\left(\frac{\beta-2x}{\alpha}\right)\,
    \Gamma\left(\frac{2x-\gamma}{\alpha}\right)\,
    \Gamma\left(\frac12+x\right)
  }{
    \Gamma\left(x\right)
  }\,
  r^{-2x}\,\txd x.
\label{rhoMellinBarnes}
\end{equation}
This integral in this expression is a contour integral involving a product of a power function and a number of gamma functions, and such integrals are known as Mellin-Barnes integrals \citep{Barnes1910, Slater1966}. This particular Mellin-Barnes integral can be written in a compact and elegant form as
\begin{equation}
\rho(r)
=
\frac{1}{\pi^{3/2}}\,
\frac{1}{\Gamma\left(\frac{\beta-2}{\alpha}\right) 
\Gamma\left(\frac{2-\gamma}{\alpha} \right)}\,
r^{-1}
H^{2,1}_{2,2}\!\left[\left.
\begin{matrix} 
(1-\frac{\beta}{\alpha},\frac{2}{\alpha}), (0,1) \\
(-\frac{\gamma}{\alpha},\frac{2}{\alpha}), (\tfrac12,1)
\end{matrix}
\,\right|
r^2
\right],
\label{rhoH}
\end{equation}
with $H^{m,n}_{p,q}$ the Fox $H$ function, generally defined as
\begin{equation}
H^{m,n}_{p,q}\!\left[\left.
\begin{matrix} 
(a_1,A_1), \ldots, (a_p,A_p) \\
(b_1,B_1), \ldots, (b_q,B_p) 
\end{matrix}
\,\right|
z
\right]
=
\frac{1}{2\pi i}
\int_{\mathcal{L}} 
\frac{\prod_{j=1}^m \Gamma(b_j+B_j s) \prod_{j=1}^n \Gamma(1-a_j-A_j s)}
{\prod_{j=m+1}^q \Gamma(1-b_j-B_j s) \prod_{j=n+1}^p \Gamma(a_j+A_j s)}\,
z^{-s}\,\txd s.
\label{definitionFoxH}
\end{equation}
The contour $\cal{L}$ in this contour integral is chosen such that it separates the poles of the two factors in the numerator. This function was introduced by \citet{Fox61} as a generalisation of the Meijer $G$ function \citep{Meijer46}, and shares many of its properties. It is a universal, analytical function that contains many special functions, including generalised Bessel functions, elliptic integrals, generalised hypergeometric functions and the Mittag-Leffler function, as special cases. Different monographs are dedicated to the identities, asymptotic properties, expansion formulae, and integral transforms of the Fox $H$ function \citep[e.g.,][]{mathai2009h}. 

\begin{figure*}
\includegraphics[width=\textwidth]{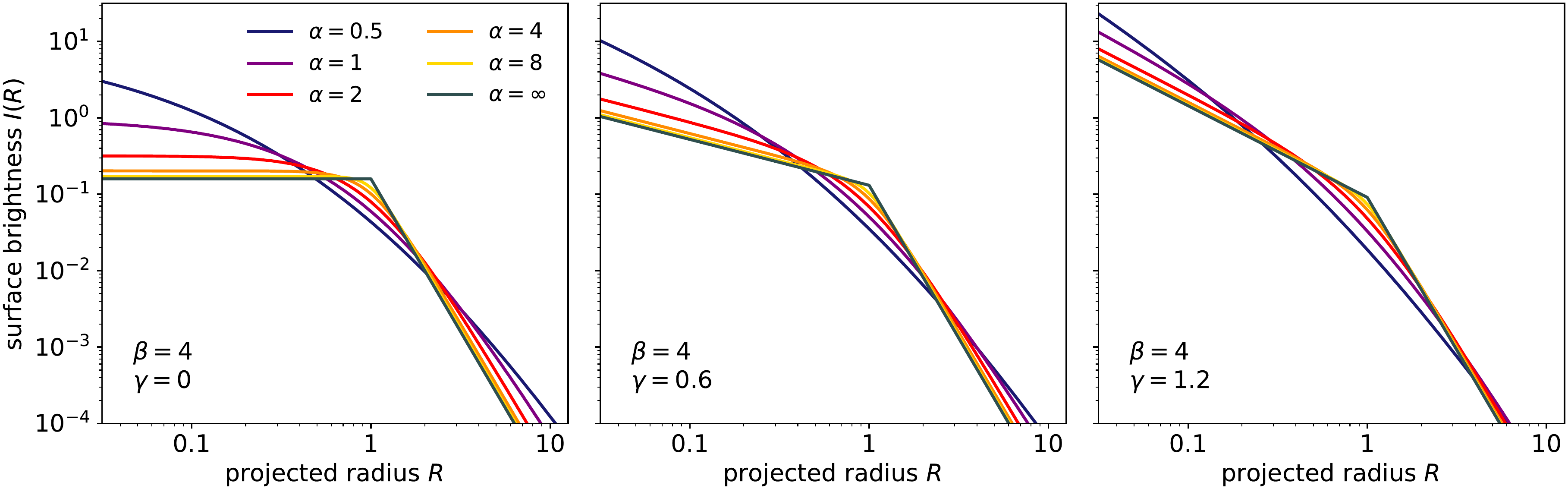}
\includegraphics[width=\textwidth]{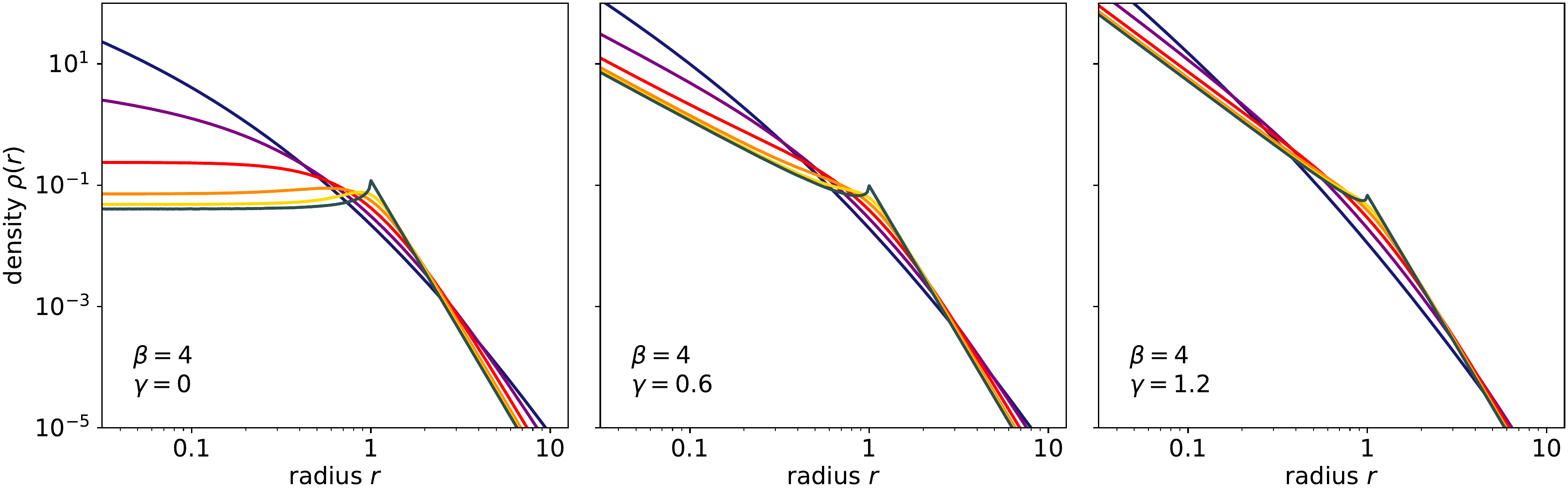}
\includegraphics[width=\textwidth]{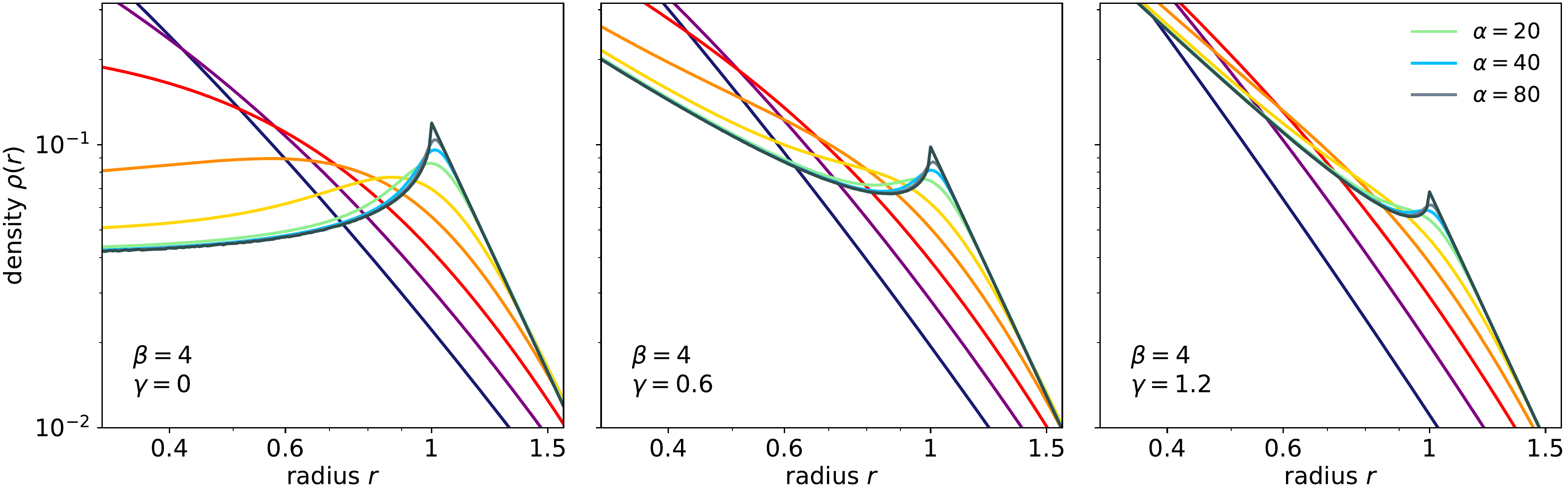}%
\caption{Surface brightness profile (top row) and density profile (second row) for the family of Nuker models. The three columns correspond to different values of $\gamma$, and different lines within each panel correspond to different values of $\alpha$. The panels on the bottom row zoom in on the density profile around the break radius $r=1$, and contain additional Nuker models with larger values of $\alpha$.}  
\label{Nuker-I-rho.fig}
\end{figure*}

The numerical evaluation of the Fox $H$ function is challenging, and general implementations of the Fox $H$ function are not (yet) available in popular numerical packages as NAG \citep{phillips1986nag}, GSL \citep{galassi2001gnu}, BOOST \citep{schaling2014boost}, or SciPy \citep{Oliphant2007}. The Fox $H$ function is a generalisation of the Meijer $G$ function, itself a generalisation of the hypergeometric function. A large variety of techniques exist for the evaluation of hypergeometric functions, including Taylor series expansions, continued fractions, quadrature methods, and more. It turns out, however, that different methods are required for different parameter and argument regimes \citep[for a review, see][]{Pearson2017}. It is therefore not surprising that a single implementation for the evaluation of the Fox $H$ function for all values of the parameters and argument is not readily available. 

The most direct approach to evaluating the Fox $H$ function is to truncate the general series expansion after a number of terms, based on some stopping criterion. This approach is generally fast, but has the drawback that the exact series expansion depends on the multiplicity of the poles of the integrand in the definition~(\ref{definitionFoxH}) of the Fox $H$ function. If all poles of the defining function are single poles, the Fox $H$ function can be expressed as a power series, whereas it becomes a logarithmic-power series if there are multiple poles \citep{mathai2009h}. Explicit power and power-logarithmic series expansions are given by \citet{KilbasSaigo99}, and summarised by \citet{2011A&A...534A..69B}. An additional drawback of this approach is that even the simple task of summing different terms in the series expansion can be accomplished in different ways, and this can have a significant impact on stability and efficiency. 

An alternative method to numerically evaluate the Fox $H$ function is based on a direct numerical integration of the contour integral in Eq.~(\ref{definitionFoxH}). This has been proven to be an interesting approach for the evaluation of general hypergeometric functions \citep[e.g.,][Sec.~5.14]{2002nrca.book.....P},. While it is generally less efficient than more standard approaches as truncated series evaluation, it has the advantage that it often is applicable to a large suite of input parameters and arguments. When adaptive Gaussian or doubly-exponential quadrature formulae are applied to the different segments of this contour, most Fox $H$ functions can be evaluated to relatively high precision. Several experimental numerical implementations for the Fox $H$ function based on this approach are available for various programming languages. Among these are relatively simple and straightforward implementations in Mathematica and Python by \citet{2012arXiv1202.2576S} and \citet{Alhennawi2016}, respectively, and a GPU-enabled Matlab implementation by \citet{2018arXiv180408101C}.

The real strength of the Fox $H$ function is its use for analytical studies, however, and as such it is gradually becoming more adopted in applied sciences, including astrophysics \citep[e.g.,][]{2006Ap&SS.305..289S, 2007BASI...35..681H, 2009ApJ...690.1280V, 2011A&A...534A..69B, 2012A&A...540A..70R, 2012A&A...546A..32R, 2012RMxAA..48..209Z, 2019A&A...626A.110B, 2019A&A...630A.113B}.

A direct check on expression (\ref{rhoH}) can be derived by calculating the total mass (or luminosity). Integrating the density profile over the entire 3D space, we obtain
\begin{equation}
M
=
\frac{2}{\sqrt\pi}\,
\frac{1}{\Gamma\left(\frac{\beta-2}{\alpha}\right) \Gamma\left(\frac{2-\gamma}{\alpha} \right)}\,
\int_0^\infty
H^{2,1}_{2,2}\!\left[\left.
      \begin{matrix} 
        (1-\frac{\beta}{\alpha},\frac{2}{\alpha}), (0,1) \\
        (-\frac{\gamma}{\alpha},\frac{2}{\alpha}), (\tfrac12,1)
      \end{matrix}
      \,\right|\,
    z
  \right]\,
  \txd z.
\end{equation} 
Setting $s=a=1$ in equation (2.8) of \citet{mathai2009h} we can immediately evaluate this integral to recover just $M=1$, as required.


\subsection{General behaviour and asymptotic expansion}

For all values of the parameters in the ranges we consider ($\alpha>0$, $\beta>2$ and $0\leqslant\gamma<2$), the Nuker model has a non-negative density. The top and middle rows of Fig.~{\ref{Nuker-I-rho.fig}} show the surface brightness profile and the density profile corresponding to Nuker models with different values of $\alpha$ and $\gamma$, and with a fixed value $\beta=4$. A full illustration of the Nuker model profiles for all possible values of the three parameters would be too much information, and we have chosen to fix $\beta$ because its value affects the shape of the surface brightness and density profiles the least. The particular value $\beta=4$ is chosen as this is the value corresponding to the \citet{1911MNRAS..71..460P} model, which is the Nuker model corresponding to $(\alpha, \beta, \gamma) = (2,4,0)$. For the Plummer model, all of the properties discussed in this paper can be calculated using elementary functions \citep{1987MNRAS.224...13D}, which gives us an interesting comparison test.

Comparing the corresponding panels on the first and second rows of Fig.~{\ref{Nuker-I-rho.fig}}, it appears that, to first order, the density profiles are similar to the surface brightness profiles, with a power-law-like behaviour at both small and large radii. We can use the explicit expression for $\rho(r)$ to investigate this in detail. Indeed, under certain conditions, always satisfied for the Fox $H$ functions considered in this paper, the asymptotic expansion of the Fox $H$ function can be calculated using the residue theorem. 

For the expansion at large radii, the dominant term is characterised by a single pole for all values of the parameters $\alpha$, $\beta$ and $\gamma$. Following the recipes outlined in \citet{KilbasSaigo99}, we find 
\begin{equation}
\rho(r)\sim 
\frac{\alpha}{2\pi^{3/2}}\,
\frac{\Gamma\left(\frac{\beta-\gamma}{\alpha}\right) \Gamma\left(\frac{\beta+1}{2}\right)}
{\Gamma\left(\frac{\beta-2}{\alpha}\right)\Gamma\left(\frac{2-\gamma}{\alpha}\right) \Gamma\left(\frac{\beta}{2}\right)}\,
r^{-\beta-1}.
\label{rhoasy}
\end{equation}
This $r^{-\beta-1}$ power-law behaviour is expected, given the $R^{-\beta}$ power-law behaviour for the surface brightness at large projected radii.

For the behaviour at small radii, we can also directly apply the recipes from \citet{KilbasSaigo99}. It takes some bookkeeping to determine the order and the multiplicity of the poles, which turn out to depend on the values of both $\gamma$ and $\alpha$. Ultimately we find after an extensive calculation
\begin{equation}
\label{rhoasy0g0}
\rho(r) \sim
\begin{cases}
\;\rho_0+\dfrac{\alpha}{4\pi^2}\,
\dfrac{\Gamma\left(\frac{\beta+3}{\alpha}\right)
\Gamma\left(\frac{\alpha-3}{\alpha}\right)}{
\Gamma\left(\frac{\beta-2}{\alpha}\right)
\Gamma\left(\frac{2}{\alpha}\right)}\,
r^2
&\qquad
{\text{if }}\gamma=0{\text{ and }}\alpha>3,
\\[2em]
\;\rho_0
+\dfrac{9}{4\pi^2}\,
\dfrac{\Gamma\left(\frac{\beta+3}{3}\right)}{\Gamma\left(\frac{\beta-2}{3}\right)\Gamma\left(\frac{2}{3}\right)}
\left[\ln\left(\dfrac{2}{r}\right)-\dfrac12-\dfrac{E+\psi\left(\frac{\beta+3}{3}\right)}{3}\right]
r^2
&\qquad
{\text{if }}\gamma=0{\text{ and }}\alpha=3,
\\[2em]
\;\rho_0+\dfrac{(\alpha-1)\,(\alpha-2)\,\beta}{4\alpha\,\pi^{3/2}}\,
\dfrac{\Gamma\left(\frac{\beta}{2}\right)
\Gamma\left(\frac{3-\alpha}{2}\right)}{
\Gamma\left(\frac{\beta-2}{\alpha}\right)
\Gamma\left(\frac{2}{\alpha}\right)
\Gamma\left(\frac{4-\alpha}{2}\right)}\,
r^{\alpha-1}
&\qquad
{\text{if }}\gamma=0{\text{ and }}2<\alpha<3,
\\[2em]
\;\rho_0-\dfrac{1}{\pi^{3/2}}\,
\dfrac{\Gamma\left(\frac{\beta+1}{2}\right)}{\Gamma\left(\frac{\beta-2}{2}\right)}\,
r^2
&\qquad
{\text{if }}\gamma=0{\text{ and }}\alpha=2,
\\[2em]
\;\rho_0-\dfrac{(\alpha-1)\,\beta}{2\alpha\,\pi^{3/2}}\,
\dfrac{\Gamma\left(\frac{\beta}{2}\right)
\Gamma\left(\frac{3-\alpha}{2}\right)}{
\Gamma\left(\frac{\beta-2}{\alpha}\right)
\Gamma\left(\frac{2}{\alpha}\right)
\Gamma\left(\frac{2-\alpha}{2}\right)}\,
r^{\alpha-1}
&\qquad
{\text{if }}\gamma=0{\text{ and }}1<\alpha<2,
\\[2em]
\;\dfrac{1}{4\pi^2}\,\dfrac{\Gamma(\beta+1)}{\Gamma(\beta-2)}\,
\left[\ln\left(\dfrac{2}{r}\right)-E-\psi(\beta+1)\right]
&\qquad
{\text{if }}\gamma=0{\text{ and }}\alpha=1,
\\[2em]
\;\dfrac{\alpha\beta}{4\pi^{3/2}}\,
\dfrac{\Gamma\left(\frac{\beta}{\alpha}\right)
\Gamma\left(\frac{1-\alpha}{2}\right)}{
\Gamma\left(\frac{\beta-2}{\alpha}\right)
\Gamma\left(\frac{2}{\alpha}\right)
\Gamma\left(\frac{2-\alpha}{2}\right)}\,r^{-1+\alpha}
&\qquad
{\text{if }}\gamma=0{\text{ and }}\alpha<1,
\\[2em]
\;\dfrac{\alpha}{2\pi^{3/2}}\,
\dfrac{\Gamma\left(\frac{\beta-\gamma}{\alpha}\right)
\Gamma\left(\frac{\gamma+1}{2}\right)}{
\Gamma\left(\frac{\beta-2}{\alpha}\right)
\Gamma\left(\frac{2-\gamma}{\alpha}\right)
\Gamma\left(\frac{\gamma}{2}\right)}\,
r^{-\gamma-1}
&\qquad
{\text{if }}\gamma>0,
\end{cases}
\end{equation}
where $E\approx0.57721566$ is Euler's constant,\footnote{Euler's constant is usually denoted by the symbol $\gamma$, but for obvious reasons we prefer to use a different symbol.} $\psi(x)$ is the digamma function, and 
\begin{equation}
\rho_0 = 
\frac{1}{\pi^2}\,
\frac{\Gamma\left(\frac{\beta+1}{\alpha}\right)
\Gamma\left(\frac{\alpha-1}{\alpha}\right)}{
\Gamma\left(\frac{\beta-2}{\alpha}\right)
\Gamma\left(\frac{\alpha+2}{\alpha}\right)}.
\label{rho0}
\end{equation}
There are several interesting aspects to this asymptotic expansion. A first observation is that all Nuker models with a cuspy surface brightness profile ($\gamma>0$) also have a cusp in their density profile. The opposite is not necessarily true, however: models with $\gamma=0$ and $\alpha<1$ do have a weak density cusp with $\rho\propto r^{-1+\alpha}$, but a finite central surface brightness. Nuker models with $\gamma=0$ and $\alpha=1$ have a logarithmical density cusp and a finite central surface brightness. 

Secondly, the asymptotic behaviour of the density for the subset of Nuker models with a density core, i.e., with $\gamma=0$ and $\alpha>1$, is still very diverse. It is particularly remarkable to look at the second term in the asymptotic expansion for these models. This term is negative for $1<\alpha\leqslant2$, indicating that the density decreases as a function of radius. For the models with $\alpha>2$, however, this second term has a positive sign, implying that the density {\em{increases}} with increasing radius in the central regions. The higher the value of $\alpha$, or equivalently, the sharper the transition between the flat inner part of the surface brightness profile and the power-law fall-off beyond the break radius, the stronger this increase: it increases from slightly stronger than linear for $\alpha\gtrsim2$ to quadratic for $\alpha>3$. This curious behaviour can easily be spotted in the left panel on the middle row of Fig.~{\ref{Nuker-I-rho.fig}}, and even better in the bottom-left panel, where we zoom in on the region around the break radius $r=1$.


\subsection{Models with an infinitely sharp break}
\label{SharpNuker.sec}

In the previous subsection, we have shown that Nuker models with $\gamma=0$ and $\alpha>2$ are characterised by a density profile that is not monotonically decreasing as a function of radius. In fact, it turns out that this is not only the case for Nuker models with $\gamma=0$. On the contrary, it is a general feature for all Nuker models, irrespective of the value of $\beta$ and $\gamma$, as long as the value of $\alpha$ is large enough.

The optimal way to illustrate this special feature is to look at the limiting subclass of Nuker models characterised by $\alpha\rightarrow\infty$. As indicated in the Introduction section, the parameter $\alpha$ characterises the sharpness of the transition between the inner and outer parts of the surface brightness profile. For small values of $\alpha$, this transition is smooth and extended, whereas the sharpness of the transition increases with increasing $\alpha$ (see top row of Fig.~{\ref{Nuker-I-rho.fig}}). In the limit $\alpha\rightarrow\infty$, the transition is infinitely sharp, and the surface brightness profile reduces to a simple broken power-law profile,
\begin{equation}
I(R)
=
\frac{(\beta-2)\,(2-\gamma)}{2\pi\,(\beta-\gamma)} \times
\begin{cases}
\;R^{-\gamma} &\qquad{\text{if }}R\leqslant1,\\[1em]
\;R^{-\beta} &\qquad{\text{if }}R\geqslant1.
\end{cases}
\label{Isharp}
\end{equation}
Apart from the `regular' Nuker models, the top panels of Fig.~{\ref{Nuker-I-rho.fig}} also show the surface brightness profile corresponding to this broken power-law model.

\begin{figure*}
\includegraphics[width=\textwidth]{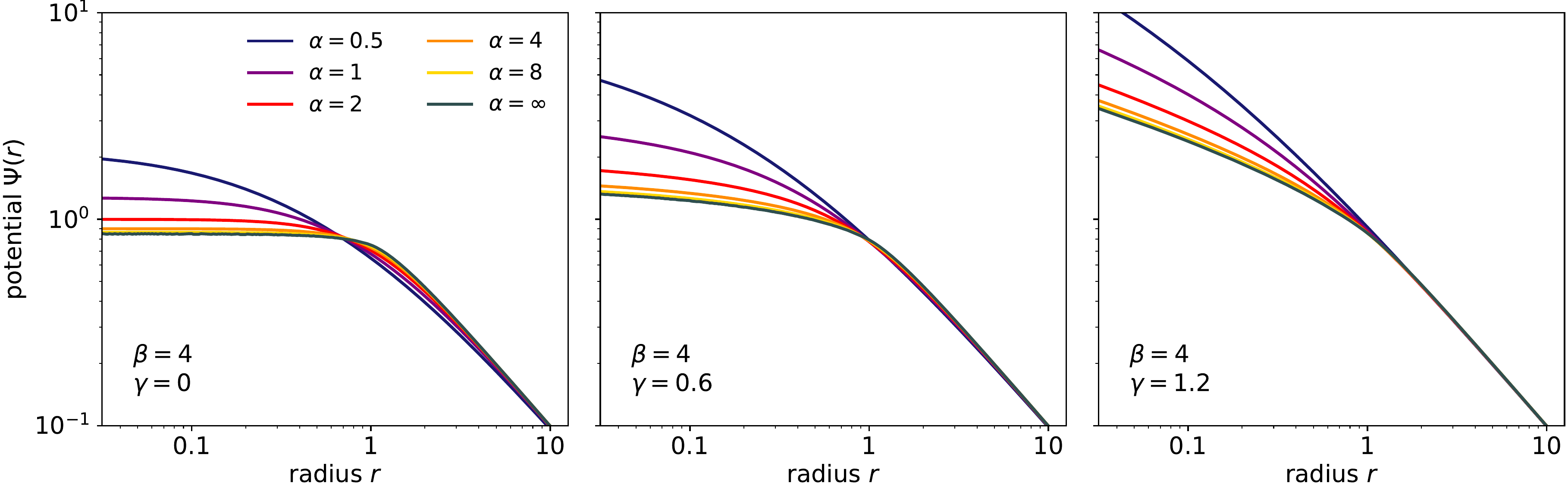}%
\caption{The gravitational potential for the family of Nuker models. The selection of models and the meaning of the different lines are as in Fig.~{\ref{Nuker-I-rho.fig}}.}  
\label{Nuker-pot.fig}
\end{figure*}

The density profile corresponding to the Nuker models with an infinitely sharp break can in principle be determined by taking the limit $\alpha\rightarrow\infty$ in expression (\ref{rhoH}). It is easier, however, to determine this limit for the explicit Mellin-Barnes expression (\ref{rhoMellinBarnes}). After some calculation, the resulting expression is a simpler Mellin-Barnes integral that can be written as a Meijer $G$ function,
\begin{equation}
\rho(r) = \frac{(\beta-2)\,(2-\gamma)}{4\pi^{3/2}}\,r^{-1}\,
G^{2,1}_{3,3}\!\left(\left.
\begin{matrix} 
1-\frac{\beta}{2}, 0, 0 \\
-\frac{\gamma}{2}, \tfrac12, --\tfrac{\beta}{2}
\end{matrix}
\,\right|\,
r^2
\right).
\label{rhosharpG}
\end{equation}
Alternatively, we can also substitute expression~(\ref{Isharp}) directly in the deprojection formula~(\ref{deprojection}). After some algebra, we find
\begin{equation}
\rho(r) = \frac{(\beta-2)\,(2-\gamma)}{2\pi^2\,(\beta-\gamma)}\,
\times
\begin{cases}
\;
\left[
\sqrt{\pi}\,\dfrac{\Gamma\left(\frac{\gamma+1}{2}\right)}{\Gamma\left(\frac{\gamma}{2}\right)}
-
\gamma\,{\cal{S}}_{\gamma}(r)
\right]
r^{-\gamma-1}
+\beta\,{\cal{S}}_{\beta}(r) \,
r^{-\beta-1}
&\qquad
{\text{if }}r\leqslant1,
\\[2em]
\;
\sqrt{\pi}\,\dfrac{\Gamma\left(\frac{\beta+1}{2}\right)}{\Gamma\left(\frac{\beta}{2}\right)}\,
r^{-\beta-1}
&\qquad
{\text{if }}r\geqslant1,
\end{cases}
\label{rhosharp}
\end{equation}
with the function ${\cal{S}}_\lambda$ defined as (see Appendix~{\ref{S.sec}}) 
\begin{equation}
{\cal{S}}_\lambda(x)
=
\int_0^{\arcsin x} \sin^\lambda\theta\,{\text{d}}\theta.
\end{equation}
The asymptotic expansion at both small and large radii is readily obtained from the asymptotic expansion of the Meijer $G$ function in Eq.~(\ref{rhosharpG}), or using the expression~(\ref{rhosharp}) and the expansion~(\ref{Sasy}) of the ${\cal{S}}_\lambda$ function. At large radii, we find a pure power-law profile, as indicated in Eq.~(\ref{rhosharp}). At small radii,
\begin{equation}
\rho(r)
\sim
\begin{cases}
\;
\dfrac{\beta-2}{\pi^2}\,
\left[\dfrac{1}{\beta+1} + \dfrac{1}{2(\beta+3)}\,r^2\right]
&\qquad
{\text{if }}\gamma=0,
\\[2em]
\;
\dfrac{(\beta-2)\,(2-\gamma)}{2\pi^{3/2}(\beta-\gamma)}\,
\dfrac{\Gamma\left(\frac{\gamma+1}{2}\right)}{\Gamma\left(\frac{\gamma}{2}\right)}\,
r^{-\gamma-1}
&\qquad
{\text{if }}\gamma>0.
\end{cases}
\end{equation}
This last expression can also be obtained by directly taking the limit $\alpha\rightarrow\infty$ in the expressions~(\ref{rhoasy0g0}) and (\ref{rho0}).

The advantage of expression~(\ref{rhosharp}) is that it is easy to investigate the behaviour of the density profile around the break radius $r=1$. For $r\gtrsim1$ we obviously have the pure $r^{-\beta-1}$ power-law fall-off that characterises the entire density profile for these models. For $r\lesssim1$, however, we obtain, thanks to Eq.~(\ref{Sasy2}),
\begin{equation}
\rho(r) 
\sim
\rho(1)
-
\frac{(\beta-2)\,(2-\gamma)}{\sqrt{2}\,\pi^2}\left(1-r\right)^{1/2}. 
\end{equation}
The coefficient of the second term in this expansion is always negative, indicating that, for $r\lesssim1$, the density profile of all Nuker models with an infinitely sharp break increases with increasing radius.

A consequence of this behaviour is that, for all values of $\beta$ and $\gamma$, there is a maximum value of $\alpha$ such that the density profile of Nuker models with $\alpha$ smaller than this value are monotonically decreasing functions of $r$, whereas models with a larger value of $\alpha$ have a non-monotonic density profile. This is clearly illustrated in the bottom row of Fig.~{\ref{Nuker-I-rho.fig}}, where we show the density for an extended set of Nuker models that also includes models with high values of $\alpha$. For models with a surface brightness core ($\gamma=0$), we already demonstrated that all models with $\alpha>2$ have an increasing density profile at small radii. The middle and right panels clearly show that also for $\gamma>0$, there is an upturn of the density profile close to the break radius if $\alpha$ sufficiently high. For $(\beta, \gamma) = (4,0.6)$ this happens for $\alpha>13.9$, for $(\beta,\gamma)=(4,1.2)$ the critical value is 31.4. The stronger the central cusp of a Nuker model, the sharper the break in the surface brightness profile needs to be to generate an upturn in the density profile.


\section{Dynamical properties of the Nuker model}
\label{Dynamics.sec}

\subsection{Gravitational potential}

Starting from the density profile (\ref{rhoH}), we can calculate the corresponding gravitational potential using the formula
\begin{equation}
\Psi(r)
=
\frac{M(r)}{r} + 4\pi \int_r^\infty \rho(s)\,s\,\txd s,
\end{equation}
where $M(r)$ is the cumulative mass distribution
\begin{equation}
M(r)
=
4\pi \int_0^r \rho(s)\,s^2\,\txd s.
\end{equation}
Starting from the Mellin-Barnes form~(\ref{rhoMellinBarnes}) for the density, we find after some algebra that both the cumulative mass profile and the gravitational potential can also be written in terms of the Fox $H$ function,
\begin{equation}
M(r)
=
\frac{2}{\sqrt\pi}\,\frac{1}{\Gamma\left(\frac{\beta-2}{\alpha}\right) \Gamma\left(\frac{2-\gamma}{\alpha} \right)}\,
r^2\,
H^{2,2}_{3,3}\!\left[\left.
\begin{matrix} 
(1-\frac{\beta}{\alpha},\frac{2}{\alpha}), (0,1), (0,1) \\
(-\frac{\gamma}{\alpha},\frac{2}{\alpha}), (\tfrac12,1), (-1,1)
\end{matrix}
\,\right|\,
r^2
\right],
\label{MH}
\end{equation}
and
\begin{equation}
\Psi(r)
=
\frac{1}{\sqrt\pi}\,\frac{1}{\Gamma\left(\frac{\beta-2}{\alpha}\right) \Gamma\left(\frac{2-\gamma}{\alpha} \right)}\,
r\,
H^{2,2}_{3,3}\!\left[\left.
\begin{matrix} 
(1-\frac{\beta}{\alpha},\frac{2}{\alpha}), (0,1), (0,1) \\
(-\frac{\gamma}{\alpha},\frac{2}{\alpha}), (-\tfrac12,1), (-1,1)
\end{matrix}
\,\right|\,
r^2
\right].
\label{PsiH}
\end{equation}
Analysing the latter expression, it is easy to demonstrate that the Nuker models are characterised by a finite potential well if $\gamma<1$,
\begin{equation}
\Psi_0 = \frac{2}{\pi}\,
\frac{\Gamma\left(\frac{\beta-1}{\alpha}\right)\,\Gamma\left(\frac{1-\gamma}{\alpha}\right)}
{\Gamma\left(\frac{\beta-2}{\alpha}\right)\,\Gamma\left(\frac{2-\gamma}{\alpha}\right)}.
\end{equation}
This result can also be obtained by plugging Eq.~(\ref{sb}) into 
\begin{equation}
\Psi_0 = -4\int_0^\infty \frac{\txd I}{\txd R}(R)\,R\,\txd R,
\end{equation}
as indicated by \citet{1991A&A...249...99C}. Models with $\gamma\geqslant1$ have a density cusp stronger than $r^{-2}$, as can be seen from Eq.~(\ref{rhoasy0g0}), and generate an infinitely deep potential well.

Using the formulae for the asymptotic expansion of the Fox $H$ function from \citet{KilbasSaigo99}, we can retrieve the explicit behaviour of $\Psi(r)$ at small and large radii. 
After some calculation, we find at small radii
\begin{equation}
\Psi(r) \sim 
\begin{cases}
\;\Psi_0-
\dfrac{\alpha}{3\pi}\,
\dfrac{\Gamma\left(\frac{\alpha-1}{\alpha}\right) \Gamma\left(\frac{\beta+1}{\alpha}\right)}
{\Gamma\left(\frac{2}{\alpha}\right)\,\Gamma\left(\frac{\beta-2}{\alpha}\right)}\,r^2
&\qquad
{\text{if }}\gamma=0{\text{ and }}\alpha>1,
\\[2em]
\;\Psi_0-\dfrac{\beta\,(\beta-1)\,(\beta-2)}{3\pi}
\left[\ln\left(\dfrac{2}{r}\right) + \dfrac56-E-\psi(\beta+1)\right]r^2
&\qquad
{\text{if }}\gamma=0{\text{ and }}\alpha=1,
\\[2em]
\;\Psi_0-\dfrac{\alpha^2}{\sqrt\pi\,(\alpha+1)\,(\alpha+2)}\,
\dfrac{\Gamma\left(\frac{1-\alpha}{2}\right)\,\Gamma\left(\frac{\alpha+\beta}{\alpha}\right)}
{\Gamma\left(\frac{2}{\alpha}\right)\,\Gamma\left(\frac{\beta-2}{\alpha}\right)\,\Gamma\left(\frac{2-\alpha}{2}\right)}\,
r^{1+\alpha}
&\qquad
{\text{if }}\gamma=0{\text{ and }}\alpha<1,
\\[2em]
\;\Psi_0-\dfrac{2}{\sqrt\pi\,(1-\gamma)}\,
\dfrac{\Gamma\left(\frac{\beta-\gamma}{\alpha}\right)\,\Gamma\left(\frac{\gamma+1}{2}\right)}
{\Gamma\left(\frac{\gamma}{2}\right)\,\Gamma\left(\frac{\beta-2}{\alpha}\right)\,\Gamma\left(\frac{2+\alpha-\gamma}{\alpha}\right)}\,
r^{1-\gamma}
&\qquad
{\text{if }}0<\gamma<1,
\\[2em]
\;
\dfrac{2}{\pi}\,
\dfrac{\Gamma\left(\frac{\beta-1}{\alpha}\right)}
{\Gamma\left(\frac{\alpha+1}{\alpha}\right)\,\Gamma\left(\frac{\beta-2}{\alpha}\right)}
\left[\ln\left(\dfrac{2}{r}\right)+1-\dfrac{E+\psi\left(\frac{\beta-1}{\alpha}\right)}{\alpha}\right]
&\qquad
{\text{if }}\gamma=1,
\\[2em]
\;\dfrac{1}{\sqrt\pi}\,
\dfrac{\Gamma\left(\frac{\beta-\gamma}{\alpha}\right)\,\Gamma\left(\frac{\gamma-1}{2}\right)}
{\Gamma\left(\frac{\gamma}{2}\right)\,\Gamma\left(\frac{\beta-2}{\alpha}\right)\,\Gamma\left(\frac{2+\alpha-\gamma}{\alpha}\right)}\,
r^{-\gamma+1}
&\qquad
{\text{if }}\gamma>1.
\end{cases}
\label{Psiasy}
\end{equation}
At large radii, we find a Keplerian decline, $\Psi(r) \sim 1/r$, as expected for a finite mass system. 

Fig.~{\ref{Nuker-pot.fig}} shows the potential for Nuker models with $\beta=4$ and with different values of $\alpha$ and $\gamma$. In all cases, the potential is a smoothly declining function of $r$. Even the models with a very sharp transition between inner and outer surface brightness profiles, which are characterised by non-monotonic density profile, have a smooth potential.


\subsection{Isotropic dynamical models}
\label{Isotropic.sec}

\begin{figure*}
\includegraphics[width=\textwidth]{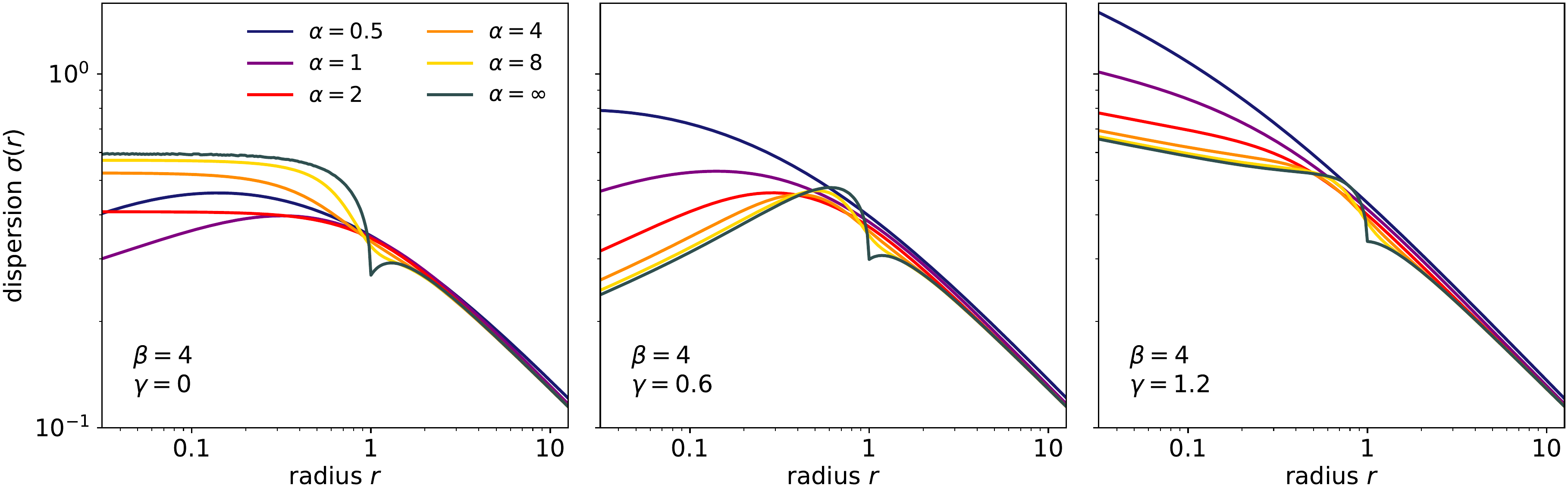}
\includegraphics[width=\textwidth]{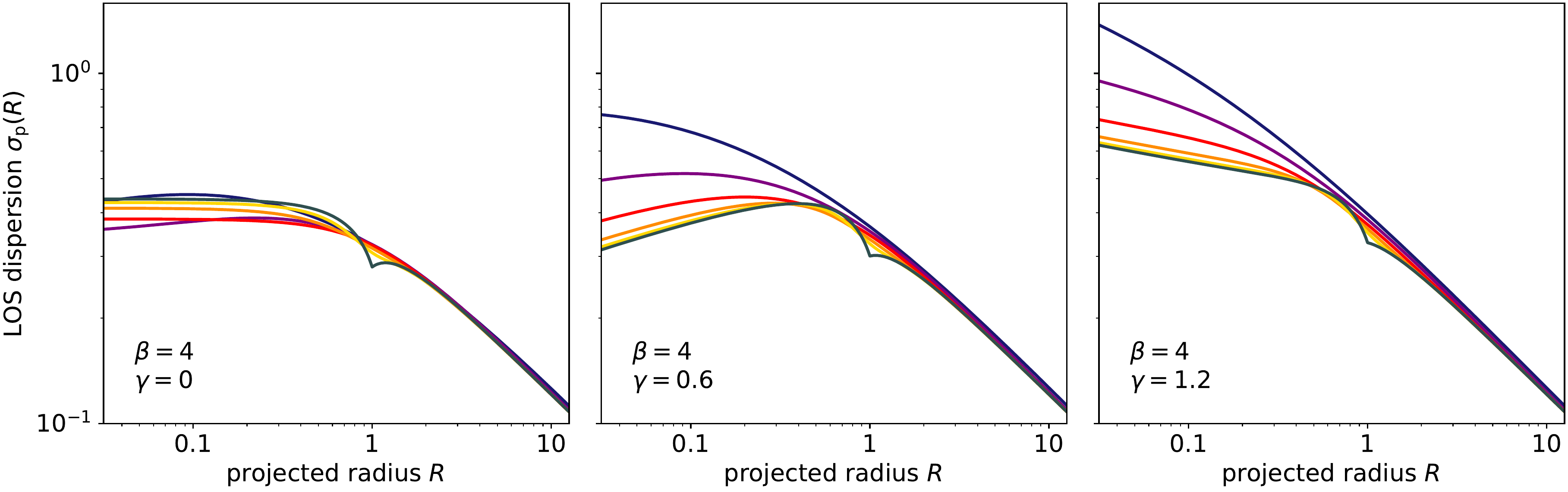}%
\caption{Intrinsic velocity dispersion profiles (top) and line-of-sight velocity dispersion profiles (bottom) for the family of isotropic Nuker models. The selection of models and the meaning of the different lines are as in Fig.~{\ref{Nuker-I-rho.fig}}.}  
\label{Nuker-sigma.fig}
\end{figure*}

For each spherical potential-density pair, we can construct a unique isotropic dynamical model that generates this density profile using standard equations of galaxy dynamics \citep[e.g.,][]{2008gady.book.....B}. The velocity dispersion profile $\sigma(r)$ of the isotropic dynamical model corresponding to a density profile $\rho(r)$ can be found using the solution of the Jeans equation,
\begin{equation}
\sigma^2(r) = \frac{1}{\rho(r)}\int_r^\infty \frac{\rho(s)\,M(s)\,\txd s}{s^2}.
\end{equation}
For the general family of Nuker models, this integral cannot be evaluated analytically. We can, however, predict how the velocity dispersion profiles will behave, based on the analysis in Appendix~C of \citet{2002A&A...386..149B}. Nuker models with $\gamma>1$, characterised by a $r^{-(\gamma+1)}$ density cusp and an infinitely deep potential well, have a velocity dispersion profile diverging as $r^{-(\gamma-1)/2}$. On the other hand, Nuker models with a weak density cusp, i.e., with $0<\gamma<1$ or $\gamma=0$ and $\alpha\leqslant1$, have a central hole in their velocity dispersion profile. Finally, the models with a finite central density, i.e., with $\gamma=0$ and $\alpha>1$ have a finite central velocity dispersion. For these models, the central dispersion $\sigma_0$ can be calculated exactly. After  some calculation and Fox $H$-function manipulation, we find
\begin{equation}
\sigma_0^2 =
\frac{4}{\alpha} 
\frac{1}
{\Gamma\left(\frac{2}{\alpha} \right) \Gamma\left(\frac{\alpha-1}{\alpha}\right) \Gamma\left(\frac{\beta+1}{\alpha}\right) \Gamma\left(\frac{\beta-2}{\alpha}\right)  
}\,
H^{3,4}_{5,5}\!\left[\left.
\begin{matrix} 
(1,\frac{2}{\alpha}),(1-\frac{\beta}{\alpha},\frac{2}{\alpha}), (\tfrac12,1), (0,1), (0,1) \\[0.2em]
(0,\frac{2}{\alpha}), (\frac{\beta}{\alpha},\frac{2}{\alpha}), (\tfrac12,1), (-1,1), (1,1)
\end{matrix}
\,\right|\,
1
\right].
\end{equation}
On the top row of Fig.~{\ref{Nuker-sigma.fig}} we show the velocity dispersion profiles of different Nuker models. The behaviour at small radii is indeed as indicated above: the left panel ($\gamma=0$) shows both dispersion profiles converging to a finite value ($\alpha>1$) and profiles with a central hole ($\alpha\leq1$), whereas the models in the central panel have velocity dispersion profiles with a central hole, and the right panels contains models with diverging dispersion profiles. Interesting is also the way the dispersion profiles change as the break become gradually sharper, for fixed values of $\beta$ and $\gamma$. For models with a smooth transition, the dispersion profiles are correspondingly smooth and featureless. When the break gets sharper, the dispersion profile shows a distinctive dive just before the break radius, which corresponds to the increase in the density. The Nuker models with an infinitely sharp break even have a sharp downward peak at $r=1$ in their dispersion profile.

Apart from the intrinsic velocity dispersion, we can also investigate the line-of-sight velocity dispersion $\sigma_{\text{p}}(R)$, which is the quantity that is observed in actual galaxies. It can be obtained by projecting the velocity dispersion along the line of sight,
\begin{equation}
\sigma_{\text{p}}^2(R) = \frac{2}{I(R)}
\int_R^\infty \frac{\rho(r)\,\sigma^2(r)\,\txd r}{\sqrt{r^2-R^2}}.
\label{sigmap2}
\end{equation}
On the bottom row of Fig.~{\ref{Nuker-sigma.fig}} we show the line-of-sight velocity dispersion profiles for the Nuker models. These profiles are similar to the intrinsic dispersion profiles, but there are some subtle differences. The main difference is the asymptotic behaviour of the velocity dispersion profile for subfamily of Nuker models with $\gamma=0$. As argued above, the behaviour of the intrinsic velocity dispersion profile of these models depends on the steepness of the break: models with $\alpha>1$ have a dispersion profile that converges to a finite value, whereas models with smaller values of $\alpha$ have a central hole in their intrinsic dispersion profile. On the contrary, the line-of-sight velocity dispersion profiles of all Nuker models with $\gamma=0$ converge to a finite value, even those models with a weak density cusp. This value can be calculated exactly, and turns out to be
\begin{equation}
\sigma_{\text{p},0}^2
=
\frac{8}{\pi \alpha^2}\,\frac{1}{\Gamma\left(\frac{2}{\alpha}\right) \Gamma\left(\frac{\beta-2}{\alpha}\right) \Gamma\left(\frac{\beta}{\alpha}\right)}\,
H^{3,4}_{5,5}\!\left[\left.
\begin{matrix} 
(\frac{\alpha-1}{\alpha},\frac{2}{\alpha}),(1-\frac{\beta}{\alpha},\frac{2}{\alpha}), (0,1), (0,1), (0,1) \\[0.2em]
(0,\frac{2}{\alpha}), (\frac{\beta-1}{\alpha},\frac{2}{\alpha}), (\tfrac12,1), (-1,1), (\tfrac12,1)
\end{matrix}
\,\right|\,
1
\right].
\end{equation}
A similar difference between intrinsic and line-of-sight velocity dispersion has been noted for other families of models with weak density cusps \citep{1993MNRAS.265..250D, 1994AJ....107..634T, 1996MNRAS.278..488Z}.

Apart from this subtle difference, the behaviour of the intrinsic and line-of-sight velocity dispersion profiles is qualitatively similar. In particular, the change in the shape of the profile for progressively sharper breaks is comparable. For models with a sharp break, the distinctive dip that is clearly visible in the intrinsic dispersion profiles is still present, but it is slightly smoothed out by the line-of-sight integration.

\begin{figure*}
\includegraphics[width=\textwidth]{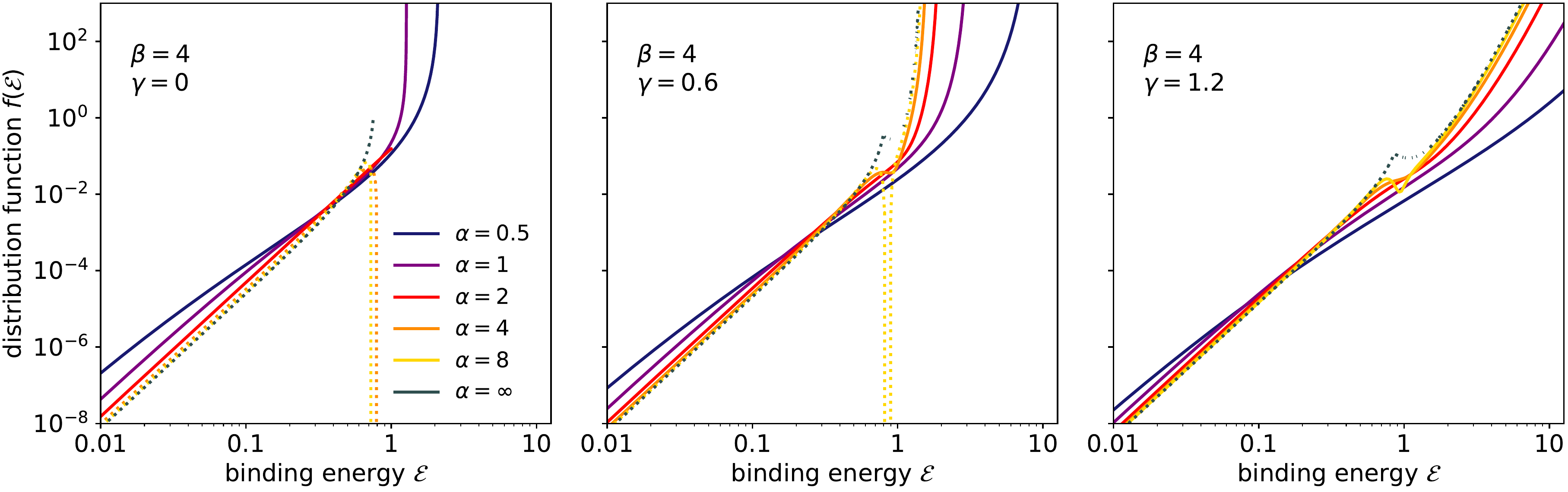}
\includegraphics[width=\textwidth]{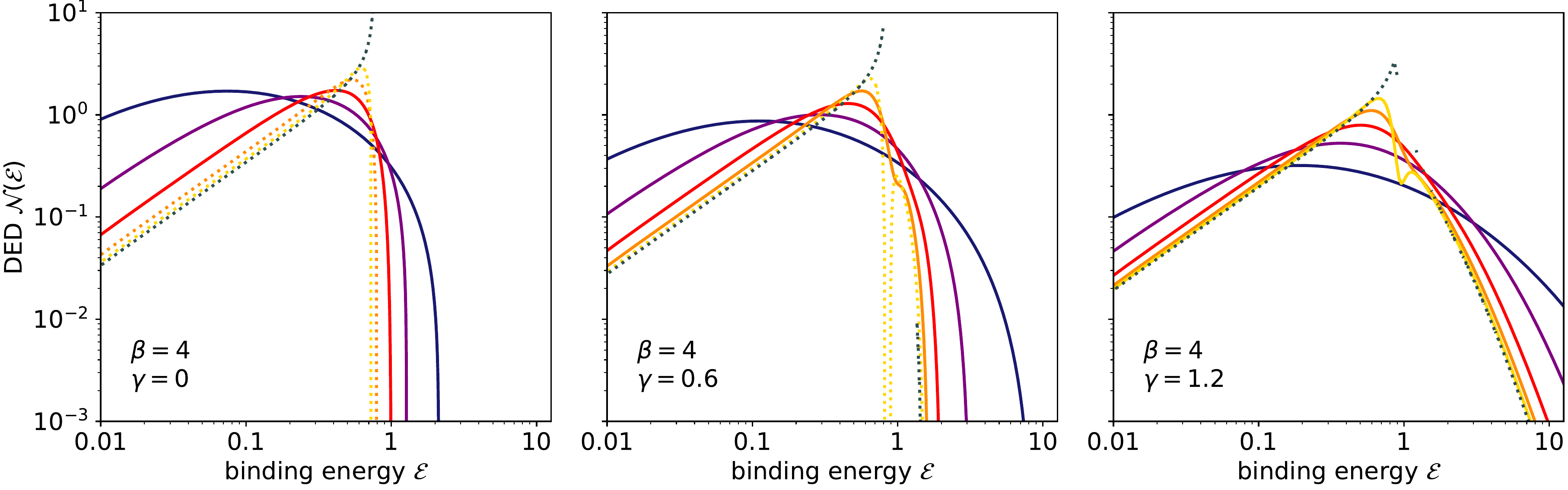}%
\caption{Distribution function (top) and differential energy distribution (bottom) for the family of isotropic Nuker models. The selection of models and the meaning of the different lines are as in Fig.~{\ref{Nuker-I-rho.fig}}. Dotted lines correspond to unphysical models, i.e., models that have a negative distribution function in some region of phase space.}  
\label{Nuker-df-ded.fig}
\end{figure*}

An important caveat on the discussion of the velocity dispersion profiles above is that they are formally derived from the solution of the Jeans equation, but it is not guaranteed a priori that they correspond to physically viable dynamical models. For a dynamical model to be physical or consistent, the distribution function $f({\boldsymbol{r}},{\boldsymbol{v}})$ must be positive over the entire phase space. It is well known that, for spherical every potential--density pair, the unique isotropic distribution function is a function of binding energy ${\cal{E}}$ only, and that it can be calculated using Eddington's formula \citep{2008gady.book.....B}. We have numerically calculated the isotropic distribution function $f({\cal{E}})$ for all the Nuker models considered before, and show the results on the top row of Fig.~{\ref{Nuker-df-ded.fig}}. The corresponding differential energy distributions ${\cal{N}}({\cal{E}})$, which describe the distribution of orbits with a given binding energy, are plotted on the bottom row. 

We have already shown that, for any values for inner and outer slopes of the surface brightness profile, there is a maximum value of $\alpha$ that guarantees that the density is a monotonically decreasing function of radius. Any models with a sharper break have an upturn in their density profile at radii $r\lesssim1$, and this weird feature translates to a downward peak or a wiggle at similar radii in the dispersion profiles. It is not surprising that these Nuker potential-density pairs do not generate physically viable isotropic dynamical models: a radially decreasing density profile is a necessary condition for the positivity of the distribution function for spherical isotropic models \citep{1992MNRAS.255..561C}. In particular, the Nuker models with an infinitely sharp break cannot be generated self-consistently by a physical isotropic distribution function. However, a monotonically decreasing density profile is not a sufficient condition for a positive distribution function, and our family of Nuker models shows that explicitly. An example is the $(\alpha, \beta, \gamma) = (8,4,0.6)$ model, characterised by the yellow line in the central columns of Figs.~{\ref{Nuker-I-rho.fig}}, {\ref{Nuker-pot.fig}}, {\ref{Nuker-sigma.fig}} and {\ref{Nuker-df-ded.fig}}. Figs.~{\ref{Nuker-I-rho.fig}} and {\ref{Nuker-pot.fig}} show that this particular model has a smoothly declining surface brightness profile, density profile and gravitational potential, without obvious signatures that would suggest a strange nature. The velocity dispersion profile of this model (top row of Fig.~{\ref{Nuker-sigma.fig}}) does show a cumbersome curvature around the break radius. Fig.~{\ref{Nuker-df-ded.fig}} shows that the distribution function and differential energy distribution of this model are negative for binding energies between ${\cal{E}}\sim0.81$ and ${\cal{E}}\sim0.88$, and hence that this model is not a physically viable isotropic model.

In general, it turns out that, for every $\beta$ and $\gamma$, the energy structure of the isotropic models varies systematically as $\alpha$ increases. All models with a soft break have positive and monotonically increasing $f(\cal{E})$, which implies that the corresponding isotropic dynamical models are not only physical, but also stable to both radial and non-radial perturbations \citep{1962spss.book.....A, 1971PhRvL..26..725D, 2008gady.book.....B}. The differential energy distributions of these Nuker models show a broad distribution of orbits over the different binding energies. 

As $\alpha$ increases, the distribution ${\cal{N}}({\cal{E}})$ becomes more peaked, and the peak gradually shifts to higher binding energies. When $\alpha$ increases even more, the distribution function starts to show a particular dip at binding energies slightly beyond the peak value. Examples of such models are the $(\alpha, \beta, \gamma) = (4,4,0.6)$ model, corresponding to the orange line in the central panels of Fig.~{\ref{Nuker-df-ded.fig}}, and the $(\alpha, \beta, \gamma) = (8,4,1.2)$ model, represented by the yellow line in the right panels. With a distribution function that is no longer a monotonically increasing function of binding energy, these models are not guaranteed to be stable against radial and non-radial perturbations anymore. For $\gamma=0.6$, the critical value of $\alpha$ that separates models with monotonic and non-monotonic distribution functions is 3.8. For $\gamma=1.2$ this value is 4.8.

Increasing $\alpha$ further strengthens the `dip' in the distribution function, until the point where it becomes negative. For $\gamma=0.6$, this point is reached for $\alpha = 6.8$, for $\gamma=1.2$ at $\alpha = 11.4$. All models with sharper break in their surface brightness distribution cannot support an isotropic distribution function. These models are indicated as dotted lines in Fig.~{\ref{Nuker-df-ded.fig}}.


\section{The S\'ersic model as a special case of the Nuker model}
\label{Sersic.sec}

The Nuker model is a very flexible model that can fit a wide range of surface brightness profiles. A special subset of Nuker models is obtained using the following recipe:
\begin{enumerate}
\item Re-introduce the scaling parameters $\Rb$ and $L$. 
\item Set $(\alpha, \beta, \gamma) = (\frac{1}{m}, \frac{1}{\epsilon m},0)$, with $m$ and $\epsilon$ positive numbers that satisfy $\epsilon m<\tfrac12$. 
\item Introduce a new length scale $\Reff = (\epsilon b)^m \Rb$, with $b$ a dimensionless number.
\item Take the limit $\epsilon\rightarrow0$.
\end{enumerate}
Applying this recipe to the Nuker surface brightness profile (\ref{sb}) gives
\begin{equation}
I(R) = 
\frac{b^{2m}}{2\pi\,m\,\Gamma(2m)}\,
\frac{L}{\Reff^2}\;
\lim_{\epsilon\rightarrow0}\,
\frac{\epsilon^{2m}\,\Gamma\left(\frac{1}{\epsilon}\right)}{\Gamma\left(\frac{1}{\epsilon}-2m\right)}\,
\left[ 1+\epsilon b\left(\frac{R}{\Reff}\right)^{1/m}\right]^{-1/\epsilon}.
\end{equation}
To evaluate this expression, we use Wendel's asymptotic relation \citep{10.2307/2304460, Qi2012},
\begin{equation}
\lim_{x\rightarrow\infty}\,x^{b-a}\,\frac{\Gamma(x+a)}{\Gamma(x+b)} = 1. \label{limgammas}
\end{equation}
This results in
\begin{equation}
I(R) = 
\frac{b^{2m}}{2\pi\,m\,\Gamma(2m)}\,
\frac{L}{\Reff^2}
\exp\left[-b\left(\frac{R}{\Reff}\right)^{1/m}\right].
\end{equation}
This is the surface brightness profile of the S\'ersic model \citep{1968adga.book.....S}, probably the most the popular model to describe the surface brightness distribution of early-type galaxies and bulges of spiral galaxies. As a result of its popularity, the properties of S\'ersic model have been examined in great detail \citep[e.g.,][]{1991A&A...249...99C, 1997A&A...321..724C, 2005PASA...22..118G}. In particular, we have used similar analytical methods as used in this paper to study the density and other intrinsic properties of the S\'ersic model in a series of papers \citep{2011A&A...525A.136B, 2011A&A...534A..69B, 2019A&A...626A.110B, 2019A&A...630A.113B}.

The analysis above shows that the S\'ersic models form a special subfamily of the general Nuker family. We can therefore study the properties of the S\'ersic model using the general expressions derived in the two previous chapters. In principle we only need to follow the four-steps recipe outlined above. If we apply this recipe to expression~(\ref{rhoH}) for the density, we find after some manipulation and application of Wendel's asymptotic relation (\ref{limgammas}),
\begin{equation}
\rho(r) 
=
\frac{b^{3m}}{\pi^{3/2}\,\Gamma(2m)}\,
\frac{L}{\Reff^3}
\left(\frac{b^m r}{\Reff}\right)^{-1}\,
\lim_{\epsilon\rightarrow0}\,
\frac{1}{\Gamma\left(\frac{1}{\epsilon}\right)}\,
H^{2,1}_{2,2}\!\left[\left.
\begin{matrix} 
(1-\frac{1}{\epsilon},2m), (0,1) \\
(0,2m), (\tfrac12,1)
\end{matrix}
\,\right|
\epsilon^{2m}\left(\frac{b^m r}{\Reff}\right)^2
\right].
\end{equation}
Using the explicit Mellin-Barnes form (\ref{definitionFoxH}) of the Fox $H$ function, this limit can be evaluated as 
\begin{equation}
\rho(r) =
\frac{b^{3m}}{\pi^{3/2}\,\Gamma(2m)}\,
\frac{L}{\Reff^3}
\left(\frac{b^m r}{\Reff}\right)^{-1}
H^{2,0}_{1,2}\!\left[\left.
\begin{matrix} 
(0,1) \\
(0,2m), (\tfrac12,1)
\end{matrix}
\,\right|
\left(\frac{b^m r}{\Reff}\right)^2
\right]
\end{equation}
in agreement with expression (17) from \citet{2019A&A...626A.110B}. Other expressions can be derived in a similar way, and it usually involves an application of Wendel's asymptotic relation. For example, from Eqs.~(\ref{rhoasy0g0}) and (\ref{rho0}) we can derive that the S\'ersic model has a finite density 
\begin{equation}
\rho_0 = \frac{b^{3m}\,\Gamma(1-m)}{\pi^2\,\Gamma(1+2m)}\,\frac{L}{\Reff^3}
\end{equation}
for $m<1$, and a power-law density cusp
\begin{equation}
\rho(r) \sim \frac{b^{3m}\,\Gamma\left(\frac{m-1}{2m}\right)}{4\pi^{3/2}\, m^2\, \Gamma(2m)\,\Gamma\left(\frac{2m-1}{2m}\right)}\,
\frac{L}{\Reff^3}\,\left(\frac{b^m r}{\Reff}\right)^{\frac{1}{m}-1}
\end{equation}
for $m>1$, in agreement with Eqs.~(19) and (20) from \citet{2019A&A...626A.110B}. In a similar way, we can derive expressions for the potential of the S\'ersic model by applying the four-steps recipe to Eqs.~(\ref{PsiH})--(\ref{Psiasy}).


\section{Discussion and summary}
\label{Discussion.sec}

\begin{figure*}
\centering
\includegraphics[width=0.75\textwidth]{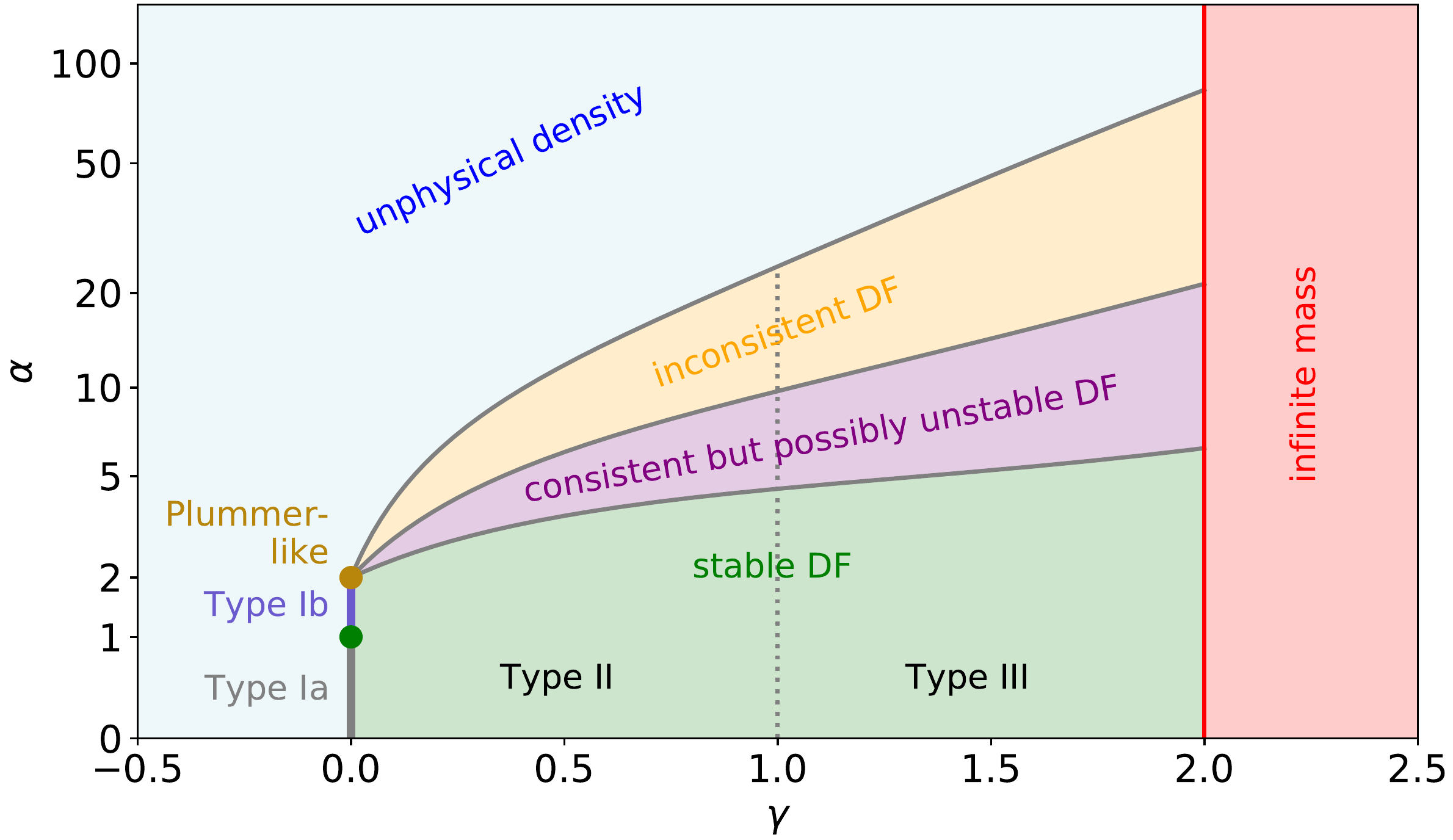}
\caption{Classification of the central structure of spherical galactic nuclei, based on the work of \citet{1994AJ....107..634T}. See text for discussion. The current figure corresponds to $\beta=4$, but it has the classification has the same qualitative behaviour for all values of $\beta>2$.}  
\label{Classification.fig}
\end{figure*}

In this paper we have constructed a three-parameter family of spherical models characterised by a simple surface brightness profile, and we have discussed the corresponding potential-density pair and the consistency and stability of the corresponding isotropic dynamical models. 

We are well aware of the obvious limitations to this set of models. One obvious limitation is that the models are spherically symmetric, and hence do not offer the rich orbital structure that more realistic triaxial models possess \citep[e.g.,][]{1991ARA&A..29..239D, 2008gady.book.....B}. There are, however, several ways to turn spherically symmetric models into flattened or triaxial models \citep{1979ApJ...232..236S, 1993ApJ...409..563S, 1989ApJ...342....1H, 1996MNRAS.281.1333D, 2009MNRAS.392.1503B}. A second limitation is that the Nuker models only provide an accurate model for the central regions of galaxies, and not for entire galaxies. Other models, such as the core-S\'ersic model proposed by \citet{2003AJ....125.2951G} have a large applicability to model the entire surface brightness profile of early-type galaxies \citep[e.g.,][]{2004AJ....127.1917T, 2006ApJS..164..334F, 2014MNRAS.444.2700D}. The Nuker model, however, remains an attractive model by virtue of its simple and still flexible form. Finally, even though the expressions (\ref{rhoH}) and (\ref{PsiH}) for density and potential are formally written as closed analytical formulae, they still involve a rather obscure special function for which a general numerical implementation is not readily available in the most popular numerical libraries. This makes a practical application of the Nuker models in direct modelling efforts less obvious. The analytical expressions do, however, allow for a detailed analytical investigation of the properties of the Nuker models, and we therefore believe that these models can be very useful for theoretical studies concerning galaxies and galactic nuclei. 

In particular, they can be used to extend and refine the classification of galactic nuclei proposed by \citet{1994AJ....107..634T}. Based on the set of $\eta$-models they presented, the authors conclude that three types of central structure are possible. The three types of central structures are denoted as flat core structure (Type~I), weak cusp structure (Type~II), and strong cusp structure (Type~III), respectively. Type~I structures, characterised by $\gamma=0$, have a finite central surface brightness, a density cusp as steep at $r^{-1}$, a finite depth of the central potential well, and an asymptotically constant line-of-sight dispersion. Type~II structures, corresponding to $0<\gamma<1$, have a density cusp ranging from $r^{-1}$ to $r^{-2}$, a finite potential well, and a central line-of-sight velocity that approaches zero near the centre. Finally, type~III structures with $1<\gamma<2$ have a density cusp stronger than $r^{-2}$, an infinitely deep potential well, and a line-of-sight velocity dispersion profile that diverges near the centre. \citet{1994AJ....107..634T} conclude that this classification does not only apply to their family of $\eta$-models, but for any spherical galaxy with an isotropic distribution function and a surface brightness profile that follows a $R^{-\gamma}$ power-law behaviour at small radii.

One limitation of the study of \citet{1994AJ....107..634T} is that the one-parameter family of models they considered is not fully representative for all central structures, even under the assumption of spherical symmetry and an isotropic dynamical structure. Our three-parameter set of models is more complete and contains a larger variance, and we believe that our models fully cover the parameter space of spherical models with an isotropic distribution function and a surface brightness profile that follows a $R^{-\gamma}$ power-law behaviour at small radii. The analysis presented in this paper has shown that $\gamma$ is the prime parameter that governs the behaviour of the Nuker models, which supports the classification of central structure based on $\gamma$ by \citet{1994AJ....107..634T}. However, we have also demonstrated that $\gamma$ alone is not sufficient to fully describe the structure of the Nuker models, and that $\alpha$ as a secondary parameter has an important role. This is particularly so for the special case of models with a flat core ($\gamma=0$), but also more generally, $\alpha$ turns out to be an important parameter for the the consistency and stability of the isotropic dynamical models. 

Based on the analysis presented in this paper, we propose to extend the classification of the central structure of spherical models as pictured in Fig.~{\ref{Classification.fig}}. The three main classes introduced by \citet{1994AJ....107..634T}, based on the value of $\gamma$, still remain. Based on the value of $\alpha$ we add some subdivision. 

For the flat core structures, we subdivide the range of models in three subclasses. Type~Ia structures correspond to $\gamma=0$ and $0<\alpha<1$. They have a cored surface brightness profile, but a weak $r^{-1+\alpha}$ cusp in their density profile. Their isotropic velocity dispersion profiles tend to zero at small radii, but their line-of-sight dispersion profile converge to a finite value. The distribution function of the isotropic dynamical models is positive and monotonic, indicating a stable isotropic dynamical structure. Type~Ib structures correspond to $\gamma=0 $ and $1<\alpha\leqslant2$. These systems are characterised by a core in both the surface brightness profile and the density. They have a finite potential density well and a non-zero central intrinsic and line-of-sight velocity dispersion. The isotropic dynamical models are consistent and stable. Finally, all models with $\gamma=0$ and $\alpha>2$ unphysical: their surface brightness distribution decreases so slowly that it can only be generated by a density profile that increases as a function of radius. This leads to negative isotropic distribution functions. The border case is formed by the one-parameter family of $(\alpha,\beta,\gamma) = (2,\beta,0)$ models, which we call the Plummer-like models. The potential-density pair of this sub-family of Nuker models simplifies substantially (Appendix~{\ref{SpecialCases.sec}}). 

For the Type~II and Type~III galaxies, a further subdivision can be made based on the monotonicity of the density profile and the consistency and stability of the isotropic distribution function. As discussed in Sec.~{\ref{SharpNuker.sec}}, there is,  for every couple $(\beta,\gamma)$, a limiting value of $\alpha$ above which the density profile shows an increase just short of the break radius. In other words, there is a limit to the sharpness of the break in the surface brightness profile if we want a physical, i.e., monotonically decreasing density profile. Furthermore, the in Sec.~{\ref{Isotropic.sec}} we showed that, for every $\beta$ and $\gamma$, the energy structure of the isotropic models varies systematically as $\alpha$ increases. Going from small to large values of $\alpha$, we first pass a region of well-behaved models that are consistent and stable. Subsequently we encounter models in which the distribution function $f({\cal{E}})$ is still positive but no longer monotonically increasing, which turns them possible unstable for radial and non-radial perturbations. Increasing $\alpha$ even more we enter into a regime where the isotropic distribution function becomes negative in some part of phase space, which implies that the isotropic models are unphysical (but anisotropic models might still exist with . Finally, we enter the previously mentioned region where the density is no longer monotonically decreasing.

\begin{acknowledgements}
The anonymous referee is acknowledged for his/her very useful and constructive comments that improved the content and presentation of this paper. MB thanks the Fund for Scientific Research Flanders (FWO Vlaanderen) and the Belgian Science Policy Office (BELSPO) for financial support.
\end{acknowledgements}

\bibliography{Nuker}

\appendix
\section{The function ${\cal{S}}_\lambda$}
\label{S.sec}

We define the function ${\cal{S}}_\lambda(x)$, with $\lambda\geqslant0$ and $0\leqslant x\leqslant1$ as
\begin{equation}
{\cal{S}}_\lambda(x)
=
\int_0^{\arcsin x} \sin^\lambda\theta\,{\text{d}}\theta.
\end{equation}
For integer values of $\lambda$, this integral is readily evaluated in terms of elementary functions, for example
\begin{gather}
{\cal{S}}_0(x)
=
\arcsin x,
\\
{\cal{S}}_1(x)
=
1-\sqrt{1-x^2},
\\
{\cal{S}}_2(x)
=
\frac12\left(\arcsin x -x\sqrt{1-x^2}\right).
\end{gather}
For non-integer values of $\lambda$ this integration is more cumbersome. A general expression can be given in terms of the hypergeometric function,
\begin{equation}
{\cal{S}}_\lambda(x)
=
\frac{\sqrt{\pi}}{2}\,
\frac{\Gamma\left(\frac{\lambda+1}{2}\right)}{\Gamma\left(\frac{\lambda+2}{2}\right)}
-
\sqrt{1-x^2}\,{}_2F_1\left(\frac12,\frac{1-\lambda}{2};\frac32;1-x^2\right).
\end{equation}
The asymptotic expansion for $x\rightarrow0$ is
\begin{equation}
{\cal{S}}_\lambda(x)
\sim
\frac{x^{\lambda+1}}{\lambda+1} + \frac{x^{\lambda+3}}{2\,(\lambda+3)}, 
\label{Sasy}
\end{equation}
whereas for $x\rightarrow1$ we find
\begin{equation}
{\cal{S}}_\lambda(x)
\sim
\frac{\sqrt{\pi}}{2}\,
\frac{\Gamma\left(\frac{\lambda+1}{2}\right)}{\Gamma\left(\frac{\lambda+2}{2}\right)}
-\sqrt2\,(1-x)^{1/2}.
\label{Sasy2}
\end{equation}

\section{Nuker models with $\alpha=2$}
\label{SpecialCases.sec}

For general values of the parameters $\alpha$, $\beta$ and $\gamma$, the density and potential of the Nuker models can be written in terms of the Fox $H$ function. For the subset of models characterised by $\alpha=2$, these formulae are reduced in complexity. Indeed, when $\alpha=2$, all components of the vectors ${\boldsymbol{A}}$ and ${\boldsymbol{B}}$ are equal to one and the Fox $H$ functions reduce to Meijer $G$ functions. Instead of the formulae (\ref{rhoH}) and (\ref{PsiH}), we obtain immediately 
\begin{gather}
\rho(r)
=
\frac{1}{\pi^{3/2}}\,
\frac{1}{\Gamma\left(\frac{\beta-2}{2}\right) \Gamma\left(\frac{2-\gamma}{2} \right)}\,r^{-1}\,
G^{2,1}_{2,2}\!\left[\left.\begin{matrix} 1-\frac{\beta}{2}, 0 \\ -\frac{\gamma}{2}, \tfrac12 \end{matrix}\,\right|\,r^2\right],
\\
\Psi(r)
=
\frac{1}{\sqrt\pi}\,
\frac{1}{\Gamma\left(\frac{\beta-2}{2}\right) \Gamma\left(\frac{2-\gamma}{2} \right)}\,
r\,
G^{2,2}_{3,3}\!\left[\left.\begin{matrix} 1-\frac{\beta}{2}, 0, 0 \\ -\frac{\gamma}{2}, -\tfrac12, -1\end{matrix}\,\right|\,r^2\right].
\end{gather}
The density can also be written in terms of the hypergeometric function,
\begin{equation}
\rho(r)
=
\frac{1}{\pi^{3/2}}\,
\frac{\Gamma\left(\frac{\beta+1}{2}\right) \Gamma\left(\frac{\beta-\gamma}{2}\right) }{\Gamma\left(\frac{\beta}{2}\right) \Gamma\left(\frac{\beta-2}{2}\right) \Gamma\left(\frac{2-\gamma}{2}\right)}\,
\left(1+r^2\right)^{-\frac{\beta+1}{2}}\,
{}_2F_1\left(\frac{\beta+1}{2}, \frac{\gamma}{2}; \frac{\beta}{2}; \frac{1}{1+r^2}\right)
\end{equation}
When $\gamma=1$, this expression for the density can be further reduced in case $\beta$ is also an integer number. For odd values of $\beta$, the density can be written purely in terms of elementary functions, for even values of $\beta$ it involves complete elliptic integrals. 

Particularly simple cases are those models with $\alpha=2$ and $\gamma=0$, a one-parameter family of Plummer-like models. In this case, we find a simple connection between the surface brightness profile and the density profile,
\begin{gather}
I(R)
=
\frac{1}{\pi}\,
\frac{\Gamma\left(\frac{\beta}{2}\right)}{\Gamma\left(\frac{\beta-2}{2}\right)}\,
\left(1+R^2\right)^{-\frac{\beta}{2}},
\label{IRa2}
\\
\rho(r)
=
\frac{1}{\pi^{3/2}}\,
\frac{\Gamma\left(\frac{\beta+1}{2}\right)}{\Gamma\left(\frac{\beta-2}{2}\right)}\,
\left(1+r^2\right)^{-\frac{\beta+1}{2}},
\end{gather}
and the gravitational potential is
\begin{equation}
\Psi(r)
=
\frac{2}{\sqrt{\pi}}\,
\frac{\Gamma\left(\frac{\beta-1}{2}\right)}{\Gamma\left(\frac{\beta-2}{2}\right)}\,
\left(1+r^2\right)^{-\frac{\beta-1}{2}}
{}_2F_1\left(\frac{\beta-1}{2},1;\frac32;\frac{r^2}{1+r^2}\right).
\end{equation}
This one-parameter family of models is not only a sub-family of the general Nuker family, but also of the class of Zhao or generalised NFW models. These models are characterised by a double power-law density profile similar to the Nuker profile, but for the spatial density rather than for the surface brightness on the plane of the sky \citep{1996MNRAS.278..488Z}. This one-parameter family of models contains a number of well-known simple models. Setting $\beta=4$, one can recognise the Plummer model \citep{1911MNRAS..71..460P, 1987MNRAS.224...13D},
\begin{gather}
I(R) = \frac{1}{\pi} \left(1+R^2\right)^{-2},
\\
\rho(r) = \frac{3}{4\pi} \left(1+r^2\right)^{-\frac{5}{2}},
\\
\Psi(r) = \frac{1}{\sqrt{1+r^2}}.
\end{gather}
For $\beta=3$ we recover the perfect sphere \citep{1985MNRAS.216..273D},
\begin{gather}
I(R) = \frac{1}{2\pi} \left(1+R^2\right)^{-\frac32},
\\
\rho(r) = \frac{1}{\pi^2} \left(1+r^2\right)^{-2},
\\
\Psi(r) = \frac{2}{\pi}\,\frac{\arctan r}{r}.
\end{gather}

\end{document}